\documentclass[aps,preprint,preprintnumbers,nofootinbib,showpacs]{revtex4}

\usepackage{amsmath}
\usepackage{amssymb}
\usepackage{mathtools}
\usepackage{dsfont}
\usepackage{color}
\usepackage{graphicx,accents}
\usepackage[normalem]{ulem}

\newcommand{\ket}[1]{\left|#1\right\rangle}
\newcommand{\pdv}[2]{\frac{\partial#1}{\partial#2}}
\newcommand{\mel}[3]{\lla{#1}|{#2}|{#3}\rra}
\def\bea{\begin{eqnarray}}
\def\eea{\end{eqnarray}}
\def\sea{\nonumber \\&&}
\def\lla{\left\langle}
\def\rra{\right\rangle}

\def\zc{\gamma}
\def\zb{\beta}

\def\ssc{\scriptscriptstyle}
\def\lsim{\mathrel{\raise.3ex\hbox{$<$\kern-.75em\lower1ex\hbox{$\sim$}}} }
\def\gsim{\mathrel{\raise.3ex\hbox{$>$\kern-.75em\lower1ex\hbox{$\sim$}}} }
\makeatletter
\DeclareRobustCommand{\cev}[1]{%
  \mathpalette\do@cev{#1}%
}
\newcommand{\do@cev}[2]{%
  \fix@cev{#1}{+}%
  \reflectbox{$\m@th#1\vec{\reflectbox{$\fix@cev{#1}{-}\m@th#1#2\fix@cev{#1}{+}$}}$}%
  \fix@cev{#1}{-}%
}
\newcommand{\fix@cev}[2]{%
  \ifx#1\displaystyle
    \mkern#23mu
  \else
    \ifx#1\textstyle
      \mkern#23mu
    \else
      \ifx#1\scriptstyle
        \mkern#22mu
      \else
        \mkern#22mu
      \fi
    \fi
  \fi
}

\makeatother
\begin{document}
%\draft
\preprint{{\vbox{\hbox{NCU-HEP-k097}
%\hbox{Aug 2022}
%\hbox{rev. Jul 2023}
\hbox{rev. Jan 2024}
\hbox{ed. Jun 2024}
}}}
\vspace*{.5in}

%\begin{frontmatter}

\title{\boldmath  On Locality of Quantum Information in the Heisenberg Picture for Arbitrary States
\vspace*{.2in}}

\author{Otto C. W. Kong}
%\ead{otto@phy.ncu.edu.tw}

\address{ Department of Physics and Center for High Energy and High Field Physics,
%Center for Mathematics and Theoretical Physics, 
National Central University, Chung-li, Taiwan 32054.  \\
Quantum Universe Center, Korea Institute for Advanced Study,
  Seoul, 02455 Republic of Korea.
\vspace*{.3in}
}

%\cortext[cor1]{Corresponding author.}

\begin{abstract}
\vspace*{.1in}
The locality issue of quantum mechanics is a key issue to a proper understanding 
of quantum physics and beyond. What has been commonly emphasized as quantum 
nonlocality has received an inspiring examination through the notion of the Heisenberg 
picture of quantum information. Deutsch and Hayden established a local description 
of quantum information in a setting of quantum information flow in a system of 
qubits.  With the introduction of a slightly modified version of what we call the 
Deutsch-Hayden matrix values of observables, together with our recently 
introduced parallel notion of the noncommutative values from a more fundamental 
perspective, we clarify all the locality issues based on such values as quantum 
information carried by local observables in any given arbitrary state of a generic
composite system.  Quantum information as the {\em `quantum' values} of 
observables gives a transparent conceptual picture of all that. Spatial locality 
for a projective measurement is also discussed. The pressing question is if and
how such information for an entangled system can be retrieved through local
processes which can only be addressed with new experimental thinking.
\\[.1in]
%\begin{keyword}
\end{abstract}

\maketitle

\section{Introduction}
The question of quantum locality, or the lack of it, has been one of great 
concern for physicists working on fundamental theories since the Bohr-Einstein 
dialogue. With the modern appreciation of the great potential in the 
applications of quantum information science, a full understanding of the 
question becomes also a practical task. In 2000, Deutsch and Hayden, looking 
at quantum information flow for systems of qubits, established `the locality 
of quantum information $from$ the Heisenberg picture' \cite{DH}. The local
Heisenberg picture description, {\em i.e.} description of the subsystems is, 
of course, a complete one. That ``a complete description of a composite system 
can always be deduced from complete descriptions of its subsystems" \cite{DH}. 
Otherwise, reduced density matrices certainly give a local description of the 
subsystems at the expense of the loss of some information about the full system.

The important result of the locality of quantum information $in$ the Heisenberg 
picture has not been as well-appreciated as it should be in the over twenty years 
after the publishing of Ref.\cite{DH}, and ``Today, most researchers in quantum 
foundations are still convinced not only that a local description of quantum 
systems has not yet been provided, but that it cannot exist." \cite{B0}. Many 
simply believe the lack of locality in quantum mechanics has been firmly 
established. However, the standard understanding of the (non)locality issue
is based heavily on keeping the Newtonian space-time picture. Our recent work 
on an alternative picture of quantum mechanics as a theory of particle dynamics 
on a noncommutative geometric (phase-)space with the position and momentum 
observables as coordinates \cite{081,078}, we believe, provide a different 
conceptual perspective on the locality issue. The perspective is exactly one 
in the Heisenberg picture, {\em i.e.} seeing the theory as, carrying  information, 
about the values of the observables. Here, the values are `quantum' in 
nature, like a kind of noncommutative number. As such observable 
has a definite state-dependent (noncommutative) value that is dictated by 
the theory of quantum mechanics, as explicitly presented in Ref.\cite{079}.
Looking at the notion of `the locality of quantum information {\em from} the 
Heisenberg picture' as presented in Refs.\cite{DH,B0,B} from the point of view 
of a fundamental theorist, we also see a few conceptual difficulties causing 
confusion. Clarifying the issues is clearly very important, and our  fundamental 
notion the values of physical quantities \cite{088} can only be better described 
in terms of mathematical systems, numbers, with a noncommutative rather 
than commutative product is exactly in line with the notion of quantum
information presented in the paper by Deutsch and Hayden \cite{DH}. It is
about the proper way to look at quantum information, at least from the 
theoretical point of view. We say {\em a piece of quantum information is 
a (noncommutative) `value' of a quantum observable}. We make an effort to 
clarify all the issues completely here, based on the statement.

It is important to note that a local description of the full quantum information 
is one thing, being able to retrieve all that information locally is another. We do 
not have, at this point, an idea about how to directly obtain a noncommutative 
value experimentally. Nor has anyone exactly demonstrated that the 
corresponding full information, of a noncommutative value or its equivalence, 
for a local observable in an entangled system can indeed be retrieved locally. 
What is important though is that there is, apparently, no good reason to see 
that as impossible. As a theorist, we are focusing here first only on the question 
of how the perspective allows us to see a different side of the story and to see 
quantum information in a different light which is at least theoretically valid. 
The experimental challenge in relation is wide open. The usual projective 
measurements certainly have hardly anything to do with retrieving the 
noncommutative values. The current study is certainly not about 
criticizing the latter and all the important results we have learned from Bell 
and others concerning that. It is more complementary to all that, providing 
alternative thinking about the nature of quantum information which has 
important implications about its handling theoretically, and very probably also 
experimentally and technologically. The actual understanding of projective 
measurements as a physical process is mostly beyond the scope of our 
analysis. We make no commitment here on the latter subject matter and 
believe our results have no necessary conflict with any particular theory of 
the kind. We also address somewhat the question of spatial locality at the 
end, which is usually phrased in terms of projective measurement. 
 We illustrate explicitly that what has been talked about as 
 a `local'  projective measurement when it is seen as a quantum dynamical, 
 {\em i.e.} unitary, process is certainly not a local one, hence the 
 noncommutative values of the local observables for the other parts of the 
 system would generally change. The same applies to its generalization as 
 any  `local'  POVM, as state projections are still involved.

Let us first clarify a background perspective. A physical theory describes
the dynamical behavior of a definite system, say a particle, or two. The dynamical
theory of quantum mechanics as such is about the kinematic set-up and the
equation of motion, as given in the Schr\"odinger or the physically equivalent
Heisenberg, picture. On the practical side, many physicists may see the corresponding
theory of measurement, as given in the projection postulate, as an integral part 
of the theory. However, how we can describe the dynamical behavior of a physical
system is one thing, how we may practically obtain information about it is another. 
It has already been appreciated that projective measurement, as described by 
the projection postulate, is not the only possible way to extract information from
a quantum system. Besides, unlike the unitary quantum evolution, the projection 
postulate in itself is not a dynamical theory, though there are available in the
literature various proposals of such a dynamical theory behind, within  the 
framework of quantum dynamics or otherwise. There are also physicists seeing
projective measurement not as a physical process. In this article, we focus only
on the dynamical theory of quantum (unitary) evolution, mostly independent
of the projective measurement. We are interested firstly in the best way one can 
describe the quantum information, for an observable at a fixed state, as given
in the mathematical dynamical quantum theory, especially the notion of locality 
in relation. Some of the questions on the practical side, in relation to projective
measurements or otherwise, will be addressed at the end. Note that the direct 
results of projective measurements are really pieces of classical information. The
complete quantum information the dynamical theory of quantum mechanics
offers is really beyond even the full statistical distribution of results of the
corresponding projective measurement on an ensemble of the same states. 
We want to emphasize though that a single projective measurement
does not provide any useful information about the quantum system, only
a statistic of such a measurement does. The latter as information for
the quantum system is certainly beyond the notion of a single commutative, 
real number, quantity. Even checking the correlation of results of projective
measurements of different parts of an entangled system has to rely on the 
kind of statistics. Those statistics are predicted by the quantum theory and 
independent of, any theory on, how the actual projection in each projective
measurement happens. However, those statistics cannot be used to provide 
complete information about the system. 

The Deutsch-Hayden {\em descriptors} as complete quantum information about 
a composite system of qubits are given as time-evolved observables, basic 
observables in terms of which all other single-qubit observables can be expressed 
in terms of. The idea that  {\em ``the `values' are matrices" \cite{DH} is an idea of 
fundamental importance}, though the phrase and its implications may easily be 
missed by readers. It shows up only once without much elaboration. The usual 
notion that the value of an observable has to be a real number is no more than a model 
assumption that enjoys great success in classical physics. When it is unquestionably 
adopted in quantum physics, we have to give up the idea of observables having 
definite values. More importantly, we cannot recover the value of an observable as 
a product of two observables from the product of values. That makes such values 
useless for checking the validity of algebraic relations among observables 
experimentally. The descriptors as matrix values can overcome the problem. 
It is not just about the basic observables in a descriptor. The approach to 
quantum information should be seen as {\em assigning a matrix value to every 
observable when a state is taken/given}, as the original authors kind of hinted at. 
For any state, the matrix values of the observables preserve among themselves 
all the algebraic relations among the observables as variables. The version of 
Deutsch-Hayden matrix value we present below makes that plainly explicit. 
Independently, we have argued,  from a different line of work, that the latter 
properties are what a proper model of values for observables should have
 \cite{081,088}. Unaware of the Deutsch-Hayden descriptors, we sought a workable 
mathematical formulation of such a notion of noncommutative values of observables 
and got the results first presented in Ref.\cite{079}. The Deutsch-Hayden matrix value
is essentially a different mathematical form of the notion. 

The Heisenberg picture focuses on using observables to describe (the evolution of) 
a quantum system. The particular dynamical evolution as quantum information flows 
in systems of qubits is of great practical importance. However, we expect our physical 
theories to give a clear picture of the complete information, as values of the basic 
observables at any fixed instance of time, which can always be identified as time $t=0$ 
independent of any particular evolution dependent on a physical Hamiltonian. Quantum 
information in a system is first about information stored in the system at a fixed instance 
of time, potentially available for future manipulations. Moreover, observables are 
dynamic variables and can have fixed values only when the state is specified. And we 
would want to be able to describe simultaneously such values for any number of states. 
Of course, we want to be able to do that for arbitrary states of any quantum system. 
Naively, the descriptor seems to have difficulty with all the issues highlighted here. 
Conceptually, as we will discuss in some detail in the next section, the observation that 
the Heisenberg picture gives a local description is an ingenious one. From the perspective 
of noncommutative values of observables, presented in any form, that is indeed not 
difficult to appreciate for a generic composite quantum system.

B\'edard looked at ``the cost" of such a complete local 
description and got the answer that ``the descriptor of a single qubit has larger 
dimensionality than the Schr\"odinger state of the whole network--or of the Universe!"
 \cite{B}. That dimensionality is certainly a lot larger. While B\'edard's analyses 
certainly help to clarify a few things about the descriptors, and some essentially 
equivalent notions of the local quantum information \cite{R}, it is easy to appreciate 
that many physicists may feel uncomfortable with the latter conclusion. We will clarify 
all that with a more direct and economic version of the Deutsch-Hayden matrix values 
as values of observables. The parallel locality property of our noncommutative values
will also be presented.

After the more general and conceptual discussions in the next section, we present
explicit illustrations of the Deutsch-Hayden matrix value of an observable in a 
language emphasizing its validity for a general quantum system independent of 
any notion of time evolution in section~\ref{sec3}. Of course, demonstrating its 
locality is a key point. The parallel illustrations will then be performed for our 
noncommutative value in section~\ref{sec4}, as inspired by the work of Deutsch 
and Hayden. The Deutsch-Hayden approach may be particularly well-adapted 
to qubit systems, especially in a practical setting of quantum information flow. 
Our noncommutative value would likely be more applicable to a general system. 
As the locality of the latter is completely transparent, the illustrations also aim at 
helping to make the notion of locality of quantum information in the Heisenberg 
picture easier to appreciate by a general reader. Some concluding remarks are put 
in the last section, including a discussion about the relevant issues of projective
measurement with an explicit illustration in the case of two entangled qubits. 
The analysis is somewhat complementary to that given in Ref.\cite{HV}. 
Examples of explicit results for the Deutsch-Hayden matrix values and our
noncommutative values of the local basic observables for the two-qubit 
system and that of our noncommutative values for a two-particle system in 
the Schr\"odinger wavefunction representation are given in the appendices.

Some readers may want to see a summary of the relevant background from
the work of Deutsch and Hayden. We see our analysis presented as
readable and understandable independent of that. Moreover, there are 
differences between the scope of our analysis and that of the original paper 
of Deutsch and Hayden \cite{DH}, and we do not agree with some of the 
opinions presented there. Having stated that we summarize the relevant 
background from the paper here. Their focus was on a matrix triple of 
time-dependent local observables for each qubit under a (Heisenberg picture) time
evolution of the quantum computation network. That is, in our notation here, 
$(\sigma_{\!\ssc 1}(t) \otimes I_r, \sigma_{\!\ssc 2}(t) \otimes I_r,\sigma_{\!\ssc 3}(t) \otimes I_r)$,
where $ I_r$ is an identity matrix of the Hilbert space of the rest of the 
system, {\em i.e.} all the other qubits. The basis for the full Hilbert space and 
each of its subspaces for the individual qubits has been chosen such that all 
$\sigma_i(0)$ are exactly the standard Pauli matrices with the state fixed 
as $\left|0\dots 0\rra$. Of course, any observable for the system at any 
later time, including the nonlocal ones, can be expressed in terms of all the 
$\sigma_i(t)$ for the qubits. The key idea is that  at time $t$ these matrices 
can  be seen as the values of the corresponding observables.  The 
state description being formally fixed cannot carry any information for the 
system. All information for each such observable can be seen as encoded in
the matrix representing it. The matrix itself hence provides a notion of
value for the observable. As the value of all observables can be obtained 
from those of the basic triples, the authors called descriptors, the latter 
contains complete information about the system. Each descriptor is only 
about local observables, hence the locality of information about the quantum 
system obtained from the Heisenberg picture of the time evolution.

\section{Locality of Quantum Information in the Heisenberg Picture}
The Heisenberg picture provides a local description of a quantum system.
That is actually easy to appreciate, especially when one draws the parallel
of the classical case from the right perspective that focuses on the similarity
between the two theories. The quantum theory is not as different from the 
classical one as one may naively think. The Schr\"odinger picture is about the 
description of the state, which is well-appreciated to be nonlocal. The state 
of a composite system of parts $A$ and $B$ may be taken as a vector in the 
product Hilbert space ${\mathcal H}_{\!\ssc A} \otimes {\mathcal H}_{\!\ssc B}$. 
While a set of basis vectors for the product space can be obtained as tensor
products among a basis vector of ${\mathcal H}_{\!\ssc A}$ and one of 
${\mathcal H}_{\!\ssc B}$, a generic vector may not be decomposed into a simple 
product of a vector of ${\mathcal H}_{\!\ssc A}$ and one of ${\mathcal H}_{\!\ssc B}$. 
Hence, one does not, in general, have a notion of the exact state for $A$ or $B$ 
for the composite, beyond the partial description in terms of reduced density 
matrices which do not carry the full information. Such a feature of entanglement 
does not exist in any classical theory. Yet, when the state is given, values for 
all observables are always determined. For the quantum case, the statement is 
still correct when the `value' or `values' of an observable is taken as the full 
statistical distribution of eigenvalue outcomes for projective measurements of 
the same state. The Heisenberg picture is, of course, about the observables. 
A state is  a point in the phase space, the description of which depends on 
a chosen system of coordinates as a frame of reference. In the classical case, 
those coordinates, the position and momentum variables for a general particle, 
for example, are the basic observables. Though the usual, real or complex number,
coordinates of the quantum phase space are not taken as (quantum) observables, 
the latter can still be considered as generic Hamiltonian functions of such 
coordinates \cite{CMP}, as in the classical case. An operator $\zb$ on the 
Hilbert space, as an observable, corresponds to the generic Hamiltonian function 
\bea\label{f}
f_{\!\ssc\zb}(z_n, \bar{z}_n)= \frac{\lla \phi |\zb | \phi \rra}{\lla \phi | \phi \rra} \;,
\eea
with $z_n$ as complex coordinates of normalized state vectors 
$\left|\phi \rra= \sum_n z_n \left|n\rra$, $\{\left|n\rra\}$ an orthonormal basis. 
The Hilbert space, or the projective Hilbert space are K\"ahler manifolds, with 
matching symplectic structures that give essentially the same Hamiltonian flows
 \cite{078}.  The set of $z_n$ has an ambiguity of a common factor as an overall 
complex phase for the vector $\left| \phi \rra$, otherwise uniquely determine 
a point in the projective Hilbert space or a fixed projection operator as a pure 
state density matrix. The set $z_n$ also serves as its homogeneous coordinates. 
The noncommutative product among observables can be obtained as \cite{CMP,079}
\bea\label{kp}
f_{\!\ssc\zb\zc}= f_{\!\ssc\zb} \star_{\!\kappa} f_{\!\ssc\zc}
= f_{\!\ssc\zb}  f_{\!\ssc\zc}   
   + \sum_n  V_{\!\ssc\zb_{n}}  V_{\!\ssc\zc_{\bar{n}}} \;,
\eea
where $V_{\!\ssc\zb_n} = \pdv{}{z_n} f_{\!\ssc\zb}$
and $V_{\!\ssc\zc_{\bar{n}}}= \pdv{}{\bar{z}_n} f_{\!\ssc\zc}$.

Exactly like the classical case, there is a set of basic observables all other observables
can be expressed in terms of. For a generic particle, the set can be given by the position
 and momentum observables. We can even think about them as noncommutative, 
operator, coordinates of the phase space \cite{078,081}. For a qubit system, the basic
observables may be conveniently taken as $\sigma_{\!\ssc 3}$ and $\sigma_{\!\ssc 1}$, 
or $\sigma_{\!\ssc 3}$ and a linear combination of $\sigma_{\!\ssc 1}$ and 
$\sigma_{\!\ssc 2}$. For a composite system $AB$, an observable for $A$ is more 
than an operator $\zb_{\!\ssc A}$ on ${\mathcal H}_{\!\ssc A}$. It is an operator 
on ${\mathcal H}_{\!\ssc A} \otimes {\mathcal H}_{\!\ssc B}$ in the form
$\zb_{\!\ssc A} \otimes I_{\!\ssc B}$ and as such one can say that it knows about 
the presence of $B$. There are nonlocal observables of $AB$ that cannot be written 
as a simple tensor product of a $\zb_{\!\ssc A}$ and a $\zc_{\!\ssc B}$. The latter 
feature is shared by the classical theory. Though the product for the 
observables for the system $AB$, as operators $\zb_{\!\ssc AB}$ or as generic 
Hamiltonian functions $f_{\!\ssc\zb_{AB}}$, is not a commutative one, local
observables of $A$ or of $B$ are well defined as $\zb_{\!\ssc A} \otimes I_{\!\ssc B}$
and $I_{\!\ssc A} \otimes \zb_{\!\ssc B}$, and as such, they form two independent 
sets. The full set of local basic observables for either part put together gives
a complete set of generators for the observable algebra of $AB$ that includes the 
nonlocal observables. That can be called the locality of observables and is what is 
behind the locality of quantum information in the Heisenberg picture, as information
contained in the values of the local basic observables. The values here, of course,
have to be representations of the complete, quantum, information, contained. In 
particular, each descriptor is essentially two matrices each can be taken as a value 
for one of the two local basic observables for a qubit \cite{B}. We call the value 
a Deutsch-Hayden matrix value, which can  be defined for all observables in any 
system. Both the Deutsch-Hayden matrix values and our noncommutative values 
for the full set of local basic observables of a system give a complete local
description of quantum information in the system, though the Deutsch-Hayden line
of analyses is typically restricted to a system of qubits.
We will show, in the analysis below, the Deutsch-Hayden matrix values for the 
local basic observables, however, contain redundant information tied to the picture 
of the latter as time-evolved observables. 

The key property of a good mathematical model of the values for observables is that 
the evaluation map, given by any fixed state, which assigns each observable its 
value has to be a homomorphism from the observable algebra to the (noncommutative) 
algebra of the values of the observables. That is to say, any algebra relation 
among observables has to be preserved by their values \cite{079}. Then, the value 
of any observable can be retrieved from the values of the local basic observables. 
That is actually beyond ``the distributions of any measurement performed on the 
whole system". Being able to predict that has been taken by many as the criteria 
for completeness \cite{B}. 

Exactly as in the classical theory, an observable as a generic Hamiltonian function
is a variable. The only information it carries is the algebraic relation between 
itself and the basic observables, which defines its explicit functional form. 
The variables cannot have fixed values until the state is fixed. Yet, even for the 
quantum case, once the state is fixed, the noncommutative values of all observables 
are fixed. However, one does not have enough information to fix the descriptors.
Again, the latter contained extra information which is irrelevant to the notion
of the quantum information in the system at a particular instance of time. The
extra information is tied to how the system evolved from the Heisenberg reference
state in the past. Hence, it is really local quantum information {\em from} the
Heisenberg picture, instead of local quantum information {\em in} the Heisenberg 
picture. We will illustrate how the Deutsch-Hayden descriptor, or our noncommutative 
values for the local basic observables, incorporated all information about the state 
into the values of the observables. Our noncommutative values do that directly, 
while the descriptor does that through a particular given time evolution. 

One may call our locality description above a weak locality. There is a stronger
requirement for a fully local description. ``Descriptions of dynamically isolated
 -- but possibly entangled -- systems $A$ and $B$ are {\em local} if that of $A$ 
is unaffected by any process system $B$ may undergo, and {\em vice versa}." \cite{B}. 
Such a process is represented by a unitary transformation in the form 
$I_{\!\ssc A}\otimes U_{\!\ssc B}$. In the Heisenberg picture description of the 
process, the state is kept unchanged while observables are modified by the
transformation. A $I_{\!\ssc A}\otimes U_{\!\ssc B}$, however, commutes with any 
$\zb_{\!\ssc A} \otimes I_{\!\ssc B}$ as a local observable of $A$ and is naturally 
unaffected by any process system $B$ may undergo. Explicitly, 
\bea
(I_{\!\ssc A}\otimes U^\dag_{\!\ssc B} ) (\zb_{\!\ssc A} \otimes I_{\!\ssc B}) 
	(I_{\!\ssc A}\otimes U_{\!\ssc B} ) = \zb_{\!\ssc A} \otimes I_{\!\ssc B} \;.
\eea
Then, any properly defined values for the local observables should respect the 
same relation. We will present the explicit Deutsch-Hayden matrix value and our 
noncommutative values in the next section, illustrating their satisfying the strong 
locality requirement too.

\section{Explicit Illustrations on the Local Deutsch-Hayden Matrix Values \label{sec3}}
The Deutsch-Hayden matrix value for an observable $\zb$ is given by
\[
\zb(t)= U^\dag(t) \zb U(t) \;,
\]
where $U(t)$ is supposed to be a fixed unitary matrix giving a particular time 
evolution of the system, {\em i.e.} it is a fixed matrix of complex numbers without 
any parameter dependence. The corresponding Schr\"odinger picture of the particular 
time evolution gives $\left|\phi(t)\rra = U(t) \left| \phi(0)\rra$. We are 
interested in $\zb$ as one of the local basic observables such as 
$\tilde{\zb}_{\!\ssc A} \equiv \zb_{\!\ssc A} \otimes I_{\!\ssc B}$ 
of a composite system $AB$. Note that $U(t)$ as a matrix on the Hilbert space 
${\mathcal H}_{\!\ssc A} \otimes {\mathcal H}_{\!\ssc B}$ is in general not 
decomposable into a tensor product, hence neither is any matrix 
$\tilde{\zb}_{\!\ssc A}(t)$. It is easy to see that any $\tilde{\zb}_{\!\ssc A}(t)$ 
is local even in the strong sense. 

When a process of the subsystem $B$ acts further on the system, the Schr\"odinger 
state would change to $\tilde{V}_{\!\ssc B} U(t) \left| \phi(0)\rra$ where
$\tilde{V}_{\!\ssc B} \equiv I_{\!\ssc A} \otimes {V}_{\!\ssc B}$ with 
${V}_{\!\ssc B}$ being the unitary matrix on ${\mathcal H}_{\!\ssc B}$ which represents
the action of the process. Switching back to the Heisenberg picture, we have 
$\tilde{\zb}_{\!\ssc A}(t)$ taken to 
$U^\dag(t)\tilde{V}_{\!\ssc B}^\dag \tilde{\zb}_{\!\ssc A} \tilde{V}_{\!\ssc B} U(t)$
hence obviously unchanged. Note that the matrix $\tilde{V}_{\!\ssc B}$ representing
the action of the process of the subsystem $B$ has the representation of 
$\tilde{V}_{\!\ssc B}(t) = U^\dag(t) \tilde{V}_{\!\ssc B} U(t)$ in the Heisenberg 
picture for the time evolution of the $U(t)$ action. And, of course
\textcolor{red}{
\[
\tilde{V}_{\!\ssc B}^\dag(t) \tilde{\zb}_{\!\ssc A}(t)  \tilde{V}_{\!\ssc B}(t)
= U^\dag(t) \tilde{V}_{\!\ssc B}^\dag  U(t)  U^\dag(t) \tilde{\zb}_{\!\ssc A} U(t)
	 U^\dag(t) \tilde{V}_{\!\ssc B}  U(t) 
= U^\dag(t) \tilde{V}_{\!\ssc B}^\dag \tilde{\zb}_{\!\ssc A} \tilde{V}_{\!\ssc B} U(t) 
=  \tilde{\zb}_{\!\ssc A}(t)  \;.
\]
}
Up to here, we only presented what has been given in Ref.\cite{B} in a more general
language, mostly for completeness. Next, we illustrate how a $\zb(t)$ naively as a
time-evolved observable serves as a value for the observable. As said, a value
of an observable must have information about the state incorporated into it. It is the
state that gives observables their values. Given a fixed basis, the Hermitian matrix 
$\zb$ gives the representation of the observable as a physical quantity. It is, of 
course, state-independent. It is a variable that carries no information about the 
system. For the $\zb(t)$ then, the information can only 
be in the $U(t)$ matrix. Actually, in the name of information flows as presented 
in the Heisenberg picture, the initial state $\left| \phi(0)\rra$ is generally taken 
as a fixed Heisenberg reference state, say the first basis vector here denoted 
as $\left|0\rra$. Then all information in the state $\left|\phi(t)\rra$ the $\zb(t)$
represents the value of is completely inside the $U(t)$ matrix. Instead of seeing
$\left|\phi(t)\rra$ as the time-evolved state, one can take it simply as the particular 
fixed state the system is in at that instance of time. Any $U(t)$ 
simply as a fixed matrix satisfying the equation $\left|\phi(t)\rra = U(t) \left|0\rra$ 
then can be used to put the information in the state into $\zb(t)$ as a matrix value 
of the observable $\zb$, at the time instance $t$. In that way, one frees $U(t)$ and 
hence $\zb(t)$ from any time evolution from the past and has direct matrix values
for all observables. More explicitly, we can simply say that with the $\left|\phi\rra$,
we pick a fixed unitary matrix $U_\phi$ satisfying $\left|\phi\rra = U_\phi \left|0\rra$ 
and use it to map each observable $\zb$ to its matrix value 
$[\zb]_\phi^{\ssc DH}=U^\dag_\phi \zb U_\phi$. The new notation emphasizes its 
nature as a value determined by the state $\left|\phi\rra$, up to the somewhat arbitrary 
but fixed choice of $U_\phi$ among solutions to the equation, without any nontrivial 
dependence on time. For any chosen  $U_\phi$, the map is an algebraic homomorphism. 
In particular $[\zb\zc]_\phi^{\ssc DH} = [\zb]_\phi^{\ssc DH} [\zc]_\phi^{\ssc DH}$ 
as simply the matrix product.  

From the above, it is easy to see the redundancy in the description. The equation 
$\left|\phi\rra = U_\phi \left|0\rra$ fixes only the first column of the matrix 
$U_\phi$. The latter is exactly the vector $\left|\phi\rra$. Any $U'_\phi= U_\phi U_{-1}$  
where $U_{-1}$ is any unitary matrix in the subspace of the full Hilbert space 
complementary to the one-dimensional subspace spanned by the vector $\left|0\rra$ 
serves the purpose equally well. One may keep $\left|\phi\rra$, and hence also 
$U_\phi$, for a generic state parametrized by as many independent real variables 
as the state admits. Say, one can that as the first column of the $U_\phi$ matrix
the complex coordinates $z_n$ of $\left|\phi\rra = \sum_n z_n \left|n\rra$ over 
the full orthonormal basis, and the rest of the columns as a fixed choice of state
vectors that, together with $\left|\phi\rra$, make up a complete orthonormal set. 
That can, of course, be done with all the vectors expressed in terms of $z_n$.
It can then be easily seen that $U_\phi$, or the matrix value $[\zb]_\phi^{\ssc DH}$,
does not involve more parameters, or more degrees of freedom, than the state vector 
$\left|\phi\rra$. If one counts also the degrees of freedom allowed in the choice 
of $U_{-1}$, one would give it a much larger number of `dimensionality' \cite{B}.
Of course, in an explicit setting of quantum information flow where one is interested
in $U(t)$ as the result of some practical processes, the story is different. In that case
$U(t)$ is given by those processes and contains information about them. It is always
possible, and in our opinion desirable, to separate the description of physical 
processes from the notion of the values of observables for a system at an instance 
of time and have a description of the latter independent of the history of the system.
 
To restrict to the particular line of analyses as presented in Refs.\cite{DH,B}, 
the full system is considered to be one of many qubits and we can regard $A$ 
as a particular qubit and $B$ a composite of the rest of the qubits. A descriptor 
for qubit $A$ may then be taken, under the new notation introduced, as 
$\{ [\tilde{\sigma}_{\!\ssc 1A}]_\phi^{\ssc DH}, [\tilde{\sigma}_{\!\ssc 3A}]_\phi^{\ssc DH} \}$, 
where  ${\sigma}_{\!\ssc 1A}= \left( \begin{array}{cc}
0 & 1 \\
1 & 0 
\end{array}  \right)$
and ${\sigma}_{\!\ssc 3A}= \left( \begin{array}{cc}
1 & 0\\
0 & -1 
\end{array}  \right)$
taken as the basic observables of qubit $A$. An even more 
explicit illustration of what we discussed above in the example of a two-qubit system, 
namely having $B$ as another qubit  is presented in Appendix A. 

\section{Explicit Illustrations on the Local Noncommutative Values \label{sec4}}
Like a Deutsch-Hayden matrix value, a noncommutative value of an observable
is an element of a noncommutative algebra as the state-dependent homomorphic
image of the observable algebra. The local basic observables form a set of 
generators for the observable algebra. In accordance, their values form a set
of generators for the algebra of the values of observables for the fixed state,
or the system at the instance. A particular representation of elements in such 
an algebra of values can be given in terms of a set of real numbers with the 
set for a product observable obtainable from a noncommutative product. For our
noncommutative value, the formalism is based on the Hamiltonian function of 
Eq.(\ref{f}) and the product given in Eq.(\ref{kp}). Given the matrix elements 
$\mel{m}{\zb}{n}$, over the basis, for a Hermitian $\zb$, the noncommutative 
value of $\zb$ for the state $\left|\phi\rra$ can be taken as 
$[\zb]_\phi = \{ f_{\!\ssc\zb}|_\phi,  V_{\!\ssc\zb_n}|_\phi \}$,
{\em i.e.} the real value of $f_{\!\ssc\zb}$ and all the complex values of
$V_{\!\ssc\zb_n}$, one for each basis vector $\left|n\rra$, all evaluated 
at the state \cite{079,093}. We have 
\bea &&
V_{\!\ssc\zb_n} =  \pdv{f_{\!\ssc\zb}}{z_n} 
 =   - f_{\!\ssc\zb} \bar{z}_n
   + \sum_m \bar{z}_m \mel{m}{\zb}{n}\;.
\label{vn}
\eea 
The product is then given by $[\zb\zc]_\phi = [\zb]_\phi \star_{\!\kappa} [\zc]_\phi$
with
\bea &&
f_{\!\ssc\zb\zc}%= f_{\!\ssc\zb} \star_{\!\kappa} f_{\!\ssc\zc}
= f_{\!\ssc\zb}  f_{\!\ssc\zc}   
   + \sum_n V_{\!\ssc\zb_n}   V_{\!\ssc\zc_{\bar{n}}} \;,
%\sea
% (\zb\zc)^m_{n} =  \sum_l (\zb)^m_{l}  (\zc)^l_{n}  \;,
\sea
V_{\!\ssc\zb\zc_n}  =   - f_{\!\ssc\zb\zc} \bar{z}_n   + \sum_m \bar{z}_m \mel{m}{\zb\zc}{n}
	=   - f_{\!\ssc\zb\zc} \bar{z}_n   + \sum_{m,l} \bar{z}_m \mel{m}{\zb}{l}\! \mel{l}{\zc}{n}\;,
\label{ncx}
\eea
where $V_{\!\ssc\zc_{\bar{n}}}=\partial_{\bar{n}} f_{\!\ssc\zc}=\overline{V}_{\!\ssc\zc_{n}}$
is just the complex conjugate of  $V_{\!\ssc\zc_{n}}$ \cite{079,093}. The
$V_{\!\ssc\zb_n}$ has only the ambiguity of an overall phase factor, from the
coordinate expressions of the state. As such, it is a common factor for all
$V_{\!\ssc\zb_n}$ of all $\zb$ in the observable algebra. All $V_{\!\ssc\zb_n}|_\phi$
have the value zero for $\left|\phi\rra$ being an eigenstate of the observable 
$\zb$. In that particular case, the noncommutative value becomes essentially 
a commutative value, {\em i.e.} its product with another noncommutative value 
is commutative, namely equals to multiplying the latter for an overall 
factor of the $f_{\!\ssc\zb}$ value, which is exactly the eigenvalue in the case. 
We have also written down an expression for noncommutative value with the state
given as a Schr\"odinger wavefunction explicitly. As the value of a wavefunction
$\phi(x_i)$ at a point of fixed values for the three $x_i$ is really a 
coordinate for the phase space, like a $z_n$, one has in the case the sequence
of $V_{\!\ssc\zb_n}$ simply replaced by the functional derivative 
$\delta_\phi f_{\!\ssc\zb}$ \cite{093}.

Locality of the noncommutative value of local observables in a composite system
is obvious when one checks. As the noncommutative value is based on
the expectation value function $f_{\!\ssc\zb}$, the value for a local observable
$[\tilde{\zb}_{\!\ssc A}]_\phi$ clearly does not change under a process on
subsystem $B$ of the composite system $AB$, given by a unitary matrix 
$I_{\!\ssc A} \otimes U_{\!\ssc B}$. In the Heisenberg picture description of 
the process, the state is untouched, and the commutativity of the observable 
$\tilde{\zb}_{\!\ssc A} \equiv {\zb}_{\!\ssc A} \otimes I_{\!\ssc B}$ and 
$I_{\!\ssc A} \otimes U_{\!\ssc B}$ says it is not changed by the transformation.
The expectation value function $f_{\!\ssc \tilde{\zb}_{\!\ssc A}}$ is invariant.
And again, the map $\zb \to [\zb]_\phi$ as an algebraic homomorphism \cite{079,CMP}
guarantees that the values of all observables, local or otherwise, are obtained
from those of the set of all local basic observables. Note that a naive 
Schr\"odinger picture thinking about the situation would seem to give a different 
conclusion. As the state is transformed, the values of the ${z}_{n}$ coordinates 
change, and so do the $V_{\!\ssc\zb_n}$ values. However, that is really a change 
in the representation of the noncommutative values rather than changes in the 
values themselves. The feature has been analyzed under quantum reference frame 
transformations in Ref.\cite{093}. An explicit representation of the noncommutative
values, just like that of the Deutsch-Hayden matrix value, is simply dependent on 
the choice of basis for the Hilbert space. One needs always to be sure to check 
changes in such values under the same representation. After all, invariance of the 
expectation value function, instead of only its value at a single point, {\em i.e.}
for a particular state, implies the invariance of all $V_{\!\ssc\zb_n}$ as
its coordinate derivatives.

Taking subsystem $A$ as a 
qubit, we have the very simple results for the values of its local basic observables 
$[\tilde{\sigma}_{\!\ssc 1A}]_\phi$ and $[\tilde{\sigma}_{\!\ssc 3A}]_\phi$. 
We have
\bea&&
f_{\!\tilde{\sigma}_{\!\ssc 1A}} 
 = \sum_n \left( \bar{z}_{{\ssc 0}n} {z}_{{\ssc 1}n} + \bar{z}_{{\ssc 1}n} {z}_{{\ssc 0}n} \right),
\sea
V_{\!\tilde{\sigma}_{\!\ssc 1A}}^{{\ssc 0}n}
 = \bar{z}_{{\ssc 1}n} - f_{\!\tilde{\sigma}_{\!\ssc 1A}} \bar{z}_{{\ssc 0}n} \;,
\qquad
V_{\!\tilde{\sigma}_{\!\ssc 1A}}^{{\ssc 1}n} 
 = \bar{z}_{{\ssc 0}n} - f_{\!\tilde{\sigma}_{\!\ssc 1A}} \bar{z}_{{\ssc 1}n} \;,
\sea
f_{\!\tilde{\sigma}_{\!\ssc 3A}} = \sum_n \left( |z_{{\ssc 0}n}|^2 - |z_{{\ssc 1}n}|^2 \right),
\sea
V_{\!\tilde{\sigma}_{\!\ssc 3A}}^{{\ssc 0}n} = \bar{z}_{{\ssc 0}n} (1- f_{\!\tilde{\sigma}_{\!\ssc 3A}})\;,
\qquad
V_{\!\tilde{\sigma}_{\!\ssc 3A}}^{{\ssc 1}n} = -\bar{z}_{{\ssc 1}n} (1+ f_{\!\tilde{\sigma}_{\!\ssc 3A}})\;,
\label{bo-nc}
\eea
where we have used
$\left|\phi\rra= \sum_n \left( {z}_{{\ssc 0}n} \left|0n\rra + {z}_{{\ssc 1}n} \left|1n\rra \right)$
and put the coordinate indices for the derivatives of $f_{\!\tilde{\sigma}_{\!\ssc 1A}}$
and $f_{\!\tilde{\sigma}_{\!\ssc 3A}}$ as superscripts for convenience. Again,
full explicit results are given for the examples of an entangled state in Appendix A.

\section{Concluding Remarks}
Unlike the notion of a local state for a subsystem in a composite system, 
local observables are always well-defined for a quantum system, with or 
without entanglement. Heisenberg picture is firstly about describing things 
with observables. To have a definite description, we need a state-specific
notion of its value, Deutsch and Hayden introduced a notion of matrix values 
for the observables, though originally cast in the language of time-evolved 
observables. We provide here a more direct description of them simply as
values for the observables at a fixed instance of time. From the perspective
of fundamental theories, we have recently introduced a more direct notion 
of noncommutative values for the observables. And, naturally, local observables 
for a subsystem are not affected by any process other subsystems may be 
subjected to. The same holds for their values, as our detailed analyses clearly 
show. The long `established' emphasis of the quantum theory being nonlocal and 
the persistence of sticking to the idea that a value has to be a real number make 
it difficult to realize the kind of local description. Despite that, Deutsch 
and Hayden, in the study of quantum information flow, saw the light. After 
all, the information contained in or carried by a quantum system is quantum 
information and the information is not quantum if each piece of it can be 
described by a real number. The Deutsch-Hayden matrix values or our form 
of the noncommutative values presented are each simply a description of the 
full quantum information contained in an observable as the mathematical model 
of a physical quantity.

From the introduction of our notion of noncommutative values, we have 
emphasized that they are elements of a noncommutative algebra, serving
as a homomorphic image of the observable algebra under an evaluation map
given by the state. That is the proper picture of what the values of observables 
should be in general. In the classical case, the observable algebra is a commutative 
one, and so is any of its algebra of values, giving the latter as essentially 
a subalgebra of the algebra of real numbers. Early in his study of quantum 
mechanics, Dirac talked about classical observables as c-number quantities and 
quantum observables as q-number ones. The observables themselves are c-number,
namely real number, and q-number variables, respectively. By definition, c-number 
variables take c-number values. In exact analog, q-number variables should take 
q-number values. What has been missing is exactly the q-numbers or a kind of 
noncommutative numbers. Deutsch and Hayden gave the first description 
of that, though not explicitly for all the observables, and without addressing the full 
algebraic properties of the set of such q-numbers. The noncommutative algebra of 
their matrix values for a fixed state is, when stripped of the irrelevant parts, simply 
a different mathematical presentation of the same algebra of q-numbers as our 
noncommutative values. In a definite practical setting of quantum information 
flow, however, the original Deutsch-Hayden formulation as time-evolved 
observables carrying extra information about the past time evolution may be  of 
other interest in applications. That extra information is an independent part, which 
is about how those q-number values have evolved dynamically.

All observables, local or nonlocal, can be expressed in terms of a number of 
local basic observables. For a generic particle without spin, they are given by 
the position and momentum observables. The q-number values for the observables, 
as the homomorphic image, of course, have all algebraic relations among the 
observables preserved among them. Hence, the local quantum information in the
full set of q-number values for the local basic observables is complete. The 
q-number values of all observables can be retrieved from them. 
  
The popular version of the story of quantum nonlocality is described with
a projective measurement of an entangled state of a two-qubit system $AB$. 
For example, we take the state 
$\left| \psi \rra = e^{\frac{-i\zeta}{2}} \sqrt{\frac{1+r}{2}} \left| 00 \rra 
	+ e^{\frac{i\zeta}{2}} \sqrt{\frac{1-r}{2}} \left| 11 \rra$ 
used in the analysis presented in Appendix A, which has an entanglement 
given by $\sqrt{1-r^2}$ for $0 \leq r < 1$. A projective measurement on 
$\tilde{\sigma}_{\!\ssc 3B}$ yielding a eigenvalue of $1$ would 
have the state collapsed to $\left| 00 \rra$. Certainly, no local observable 
for the qubit $A$ can have a sensible identical value for the states 
$\left| \psi \rra$ and $\left| 00 \rra$. However, one needs to look at the 
situation more carefully. Some physicists do not consider the `wavefunction 
collapse' resulting from the measurement as a physical process within the 
scope of the description of quantum mechanics itself. Otherwise, we believe the 
decoherence theory \cite{dc,dc2} provides quite a successful description of that, 
at least in principle. The description involves, actually unavoidable, couplings 
of the system to the measuring apparatus and the environment. One would expect 
the latter to couple to both $A$ and $B$. The results accessible are 
only given in terms of reduced density matrices. Hewitt-Horsman and Vedral,
in their interesting effort to develop the ``Deutsch-Hayden approach" \cite{HV}
have addressed the projective measurement. We present a somewhat complementary
analysis here for further clarification of the, arguably, locality issue 
receiving the most attention. Hewitt-Horsman and Vedral consider a CNOT gate
as a model for the ``measurement-type interactions" on a qubit. More exactly,
the qubit to be measured is taken as the control and the target qubit serves
as the `apparatus/pointer'. Starting from an initial pointer state $\ket{0}$
coupling to a qubit to be measured, say the $B$ qubit of our system $AB$ in
the above state, the CNOT gate turns the state for, here, the system $ABC$ 
into one with a perfect correlation between qubits $B$ and $C$. Note however 
that a fixed pointer output of either $\ket{0}$ and $\ket{1}$, the so-called
wavefunction collapse still has not happened. Explicitly, we have
\bea\label{cnot}
\ket{\psi}_{\!\ssc AB}\otimes \ket{0}_{\!\ssc C}
\rightarrow e^{\frac{-i\zeta}{2}} \sqrt{\frac{1+r}{2}} \left| 000 \rra_{\!\ssc ABC} 
	+ e^{\frac{i\zeta}{2}} \sqrt{\frac{1-r}{2}} \left| 111 \rra_{\!\ssc ABC} \;.
\eea
The decoherence theory further invokes coupling to the many more degrees 
of freedom of the `environment', the exact initial state of which cannot be 
practically determined. The system then evolves into a macro state that has 
the subsystem $ABC$ `collapsed' to either $\ket{000}$ and $\ket{111}$, or 
more exactly the reduced density matrix of $\ket{000}\!\lla {000} \right|$ or 
$\ket{111}\!\lla {111} \right|$. The undetermined exact microstate generally 
does not stabilize as the macrostate. The macrostate outcome is supposed 
to be quantum mechanically determined, depending on the practically
inaccessible initial (micro)state of the `environment' in line with the Born 
probability expectations. Even the exact microstate of the complete system 
at any instance is supposed to be quantum mechanically determined. Short 
of going through the exact description of all such details, we want to point 
out a few things that are important to the understanding of the locality issues 
involved. First of all, concerning the strong notion of locality discussed, the 
CNOT gate as a quantum process for two qubits is not a local one. Similarly, 
the unitary evolution given by (\ref{cnot}) is not local. Explicitly, it cannot be
written in the form $I_{\!\ssc A}\otimes U_{\!\ssc BC}$ not to say  
$I_{\!\ssc A}\otimes U_{\!\ssc B}\otimes U_{\!\ssc C}$. The further part
of quantum evolution from a complete decoherence theory picture would very
unlikely be in the form of $I_{\!\ssc A}\otimes U$ either. One way or another,
a projective measurement cannot be a local process on the subsystem being
measured. More directly, the `collapse' taking the state $\left| \psi \rra$ 
to $\left| 00 \rra$ can only be implemented within the system by the unitary 
transformation $U_q^{-1}$ of the $U_q$ given in the appendix, {\em i.e.} 
$\left| 00 \rra = U_q^{-1} \left| \psi \rra$. But $U_q$, or $U_q^{-1}$, is 
not a local transformation of the form $I_{\!\ssc A} \otimes U_{\!\ssc B}$. 
We have been naively thinking about such a projective measurement to be 
a local process. That actually cannot be justified. Then, the process of
measuring subsystem $B$ causing changes to the values of local observables 
for subsystem $A$ does not violate the locality of those observables or the 
locality of quantum information in the Heisenberg picture. 

There is another notion of locality that seems to dictate the projective 
measurement to be local, namely spatial locality. One considers the case 
that $A$ and $B$ are `spatially well separated', say with spacelike separation 
in the Minkowski spacetime picture. Then measuring $B$ cannot result in 
changes in $A$ without having action-at-a-distance violating special relativity. 
The q-number value picture, however, has a deep implication that suggests 
a completely different way to look at the issue. The idea of observables 
taking values in a noncommutative algebra applying to the position 
observables gives us a new perspective on the notion of a point in 
space or an event in spacetime. Such points each described by such the 
q-number position (coordinate) values cannot be a point in the c-number
space. The classical Newtonian space and Minkowski spacetime models 
should be replaced by some noncommutative geometric models
 \cite{078,081}. Questions about spatial locality have to be reformulated 
accordingly. At this point, it is still an open question. 

We want to emphasize that the description concerning projective measurements
as given above is only an attempt to clarify some of the related questions. Our
main analysis and results are independent of that. Again, how we are to 
understand the physics of projective measurements is one thing, a good 
theoretical picture of the quantum information given by the dynamical theory 
of quantum mechanics, independent of the latter, is another. We cannot rule out
the possibility of having some other dynamical theory, local or nonlocal in their
appropriate senses, giving a better description of the physics of projective 
measurements. Nor should we rule out the possibility of having some other
ingenious experimental designs to retrieve or manipulate the quantum 
information in ways beyond the limit of projective measurements.

The whole dynamical theory of quantum mechanics can be described in the 
Heisenberg picture, that is to say, by using only the observables. With the notion
of noncommutative values for the observables, the full set of local basic 
observables can be seen as a system of noncommutative coordinates for the 
quantum phase space \cite{081,078}. From that perspective, the Schr\"odinger
picture is the c-number(real/complex number) coordinate picture while the 
Heisenberg picture is the q-number (noncommutative) coordinate picture. As 
Timpson \cite{T} discussed in the language of the Deutsch-Hayden descriptors, 
such noncommutative values of the local noncommutative coordinates can be 
seen to give descriptions of the exact local states, through values of 
their q-number coordinates, even for a composite system with entanglement. 
The full picture  describes the theory as simply 
a noncommutative version of its classical analog. Quantum reality is then 
simply a noncommutative reality \cite{088}, not necessarily practically
inaccessible or having unpalatable features as Timpson stated. It has been
discussed in Ref.\cite{088} how the real number readings we get out of our 
measuring equipment are only the results of the real number we put in to 
calibrate the scales for displaying the outputs. Otherwise, nothing in 
Nature says physical quantities have real number values. Nothing says 
that the information we have obtained, say what the pointer on the scale 
indicates, is a piece of classical information, beyond the case that the 
idea may work well enough for the familiar particular settings. In the age 
of quantum information science, it is time we think about dealing with the 
full information in a quantum observable and general quantum information 
directly. In the future, we may even find ways to experimentally determine 
the noncommutative value or q-number value of an observable directly in 
one piece.  

Finally, let us further clarify a few points against plausible confusion for the 
benefit of the readers. Firstly, we are not proposing any new theory or new 
model to describe physics. The work is theoretical, analyzing exactly 
the logical and mathematical implications of the dynamical theory of quantum 
mechanics. We are not challenging the correctness of Bell's theorem so long 
as the results of projective measurements and theories of local hidden variables, 
as c-number valued variables, are concerned. Our analysis has hardly
touched on the kind of measurements or much of any issues about practical
measurements. Neither are we questioning the nonlocal character of an 
entangled state within the Schr\"odinger picture, or more precisely so long
as the usual description of the state in terms of a commutative/c-number
geometric picture of the phase space is concerned. We have demonstrated 
that the dynamical theory of quantum mechanics, independent of any
theory of projective measurements, essentially predicts theoretical
information about an observable for a definitive state richer than what
is encoded in the full statistical distribution of the corresponding projective
measurement \cite{088,079,081}. That full theoretical information is an 
element of a noncommutative algebra, of the noncommutative values.
One can think about them as exact q-numbers. We see that as what quantum 
information is about. That quantum information for a full set of basic 
observables, which can be naturally identified among the local observables 
for a composite system, is complete in the sense that the corresponding 
information for all observables can be mathematically determined from it. 
Such local quantum information for a part of a composite system is not 
affected by any physical process local to other parts of the system, even 
for entangled states. That is our clarification of the full local picture based
on describing everything in terms of (values of) observables, which is
simply an exact analog of the classical case. We call that the Heisenberg
picture, generalizing from its usual discussion involving a time evolution.

An important question, which we have addressed before \cite{088,079,081},
is the practical accessibility of the full quantum information in a noncommutative 
value. In terms of its representation as a sequence of complex numbers, which 
is finite for a system with a finite-dimensional Hilbert space, the 
noncommutative value can be determined up to any required precision in 
principle. In the interest of the current analysis, it suffices to say that 
those involve information about the state description which is hence 
not local. Extracting quantum information is certainly not simply about 
making a few projective measurements. The notion of noncommutative 
values is about treating quantum information directly as quantum
information. It clearly indicates that there is information in the system
beyond the access of projective measurements, and the complete 
information, including the part that is nonlocal in the conventional state 
picture, is encoded in those values of the basic local observables.  
How to practically retrieve that full information is a challenge to 
experimentalists and practical quantum information scientists to 
whom the author as a fundamental theorist can only appeal. 

Another question some readers may have in mind is temporal locality.  
So long as 'nonrelativistic' quantum dynamics is concerned, time is 
not an observable and should even not be seen as a kinematic 
variable. It is only the (real) parameter describing the Hamiltonian
evolution, mathematically as a one-parameter group of symmetry 
transformation on the phase space, which is exactly the unitary 
one-parameter group of unitary transformations on the Hilbert 
space \cite{095}. As such, it does not bring in extra locality issues. 
If a unitary transformation may change the noncommutative values 
of a local variable, of course, depends on whether it is local to other 
parts of the system, as discussed above. The situation in 'relativistic'
quantum dynamics is more of a tricky story. For the standard theory
as available in textbooks, time is still just that evolution parameter.
and nothing changes. However, we see the theory as not truly
Lorentz covariant. A physical Lorentz covariant picture should have
spacetime position observables as components of a Minkowski
four-vector. We have formulated exactly such a theory \cite{087} 
and addressed the corresponding picture of its Hamiltonian 
dynamics and the key differences with the standard theory 
 \cite{087,096}. The theory is essentially an exact $1+3$ analog
of the standard  'nonrelativistic' theory with noncommuting 
four-vector position and momentum operators and a Minkowski
metric operator on the vector space of states for a particle. Locality
issues and noncommutative values concerning the time part are
then simply an exact parallel of what we have here. It is important
to note that the literature on the topic of temporal correlation and
locality is (see for example Ref.\cite{tl}) mostly not about the basic 
quantum dynamics of (pseudo-)unitary evolution, but rather about 
some theories extending that basic dynamics which essentially 
include a theory about projective measurements. We have refrained
from going in that direction and stayed from committing to any
theory of the kind. Again, our philosophy is that no matter how
important and useful projective measurements or POVM have
been, it is about a kind of process the actual physics of which  we 
do not quite have a complete understanding. How the theory of 
quantum mechanics describes nature is one thing. How a certain 
kind of process extracts information the theory predicts is another,
and the physical understanding of those processes still is somewhat 
something different. Quantum mechanics ascribes information 
to an observable part of which is not accessible to the kind of 
measurements, or any measurement with results depending only 
on those `probabilities' concerning the eigenstates 
of the particular observable measured. The full information is 
experimentally accessible in principle, though obtaining any such 
noncommutative value directly is a huge challenge. We cannot
prove that it would not turn out to be impossible, though we do 
not see any  reason to be that pessimistic at this point. We are 
interested in that full information here, hence not on the
projective measurements.

Our last word: Despite our personal inclination, logically, there 
are two ways to look at the locality question physically, 
depending on if the nonlocal information in the Schr\"odinger
picture encoded in the noncommutative values of the local
observables is indeed locally experimentally retrievable. If yes,
complete quantum information could be retrieved locally
and the nonlocal feature  in the Schr\"odinger picture is simply an 
artifact of the description of states not in terms of physical values
of observables. We can, and in that case definitely should, describe
the states using the noncommutative values of their local basic 
observables. Otherwise, one would have to admit the strange
conclusion that the noncommutative values of the local 
observables are nonlocal in themselves!

\section*{Appendix A: Deutsch-Hayden Matrix Values and Noncommutative Values of 
Local Basic Observables for the Two-qubit System}
In the two-qubit system, a generic normalized state vector can be written, 
up to an overall phase factor, as
\bea
\left| \Psi \rra = z_{00}\left| 00 \rra + z_{01} \left| 01 \rra
               + z_{10} \left| 10 \rra + z_{11} \left| 11 \rra \;,
\eea
with the complex coordinates $z_{jk}$ given by
\bea &&
z_{00}= e^{\frac{-i\zeta}{2}} q_+ c_{\!\ssc A} c_{\!\ssc B} + e^{\frac{i\zeta}{2}}q_- \bar{s}_{\!\ssc A}\bar{s}_{\!\ssc B} \;,
\sea
z_{01}= e^{\frac{-i\zeta}{2}} q_+ c_{\!\ssc A} s_{\!\ssc B} - e^{\frac{i\zeta}{2}}q_- \bar{s}_{\!\ssc A} \bar{c}_{\!\ssc B} \;,
\sea
z_{10}= e^{\frac{-i\zeta}{2}} q_+ s_{\!\ssc A} c_{\!\ssc B} - e^{\frac{i\zeta}{2}}q_- \bar{c}_{\!\ssc A} \bar{s}_{\!\ssc B} \;, 
\sea
z_{11}= e^{\frac{-i\zeta}{2}} q_+ s_{\!\ssc A} s_{\!\ssc B} + e^{\frac{i\zeta}{2}}q_- \bar{c}_{\!\ssc A} \bar{c}_{\!\ssc B}  \;, 
\eea
where $q_+=\sqrt{\frac{1+r}{2}}$, $q_-=\sqrt{\frac{1-r}{2}}$, 
$c_{\!\ssc A}= \cos(\frac{\theta_{\!\ssc A}}{2}) e^{\frac{-i\psi_{\!\ssc A}}{2}}$,
$s_{\!\ssc A}= \sin(\frac{\theta_{\!\ssc A}}{2}) e^{\frac{i\psi_{\!\ssc A}}{2}}$,
$c_{\!\ssc B}= \cos(\frac{\theta_{\!\ssc B}}{2}) e^{\frac{-i\psi_{\!\ssc B}}{2}}$,
$s_{\!\ssc B}= \sin(\frac{\theta_{\!\ssc B}}{2}) e^{\frac{i\psi_{\!\ssc B}}{2}}$,
$0 \leq r \leq 1$, $0 \leq \theta_{\!\ssc A}, \theta_{\!\ssc B} \leq \pi$,
$0 \leq \zeta, \psi_{\!\ssc A}, \psi_{\!\ssc B} < 2\pi$. 
The six real parameters 
$\{ r, \zeta, \theta_{\!\ssc A}, \theta_{\!\ssc B}, \psi_{\!\ssc A}, \psi_{\!\ssc B} \}$
uniquely specify a physical state as a pure state density matrix and a point in the 
projective Hilbert space. The value of $r$ completely characterizes the entanglement 
as $\sqrt{1-r^2}$. At $r=1$, the $\left| \Psi \rra$ reduces the general product state
$\left( c_{\!\ssc A} \left| 0 \rra + s_{\!\ssc A} \left| 1 \rra \right) \otimes
	\left( c_{\!\ssc B} \left| 0 \rra + s_{\!\ssc B} \left| 1 \rra \right)$.
%(see, for example, Ref.\cite{092}). 
For the descriptor as
$\{ [\tilde{\sigma}_{\!\ssc 1A}]_\phi^{\ssc DH}, [\tilde{\sigma}_{\!\ssc 3A}]_\phi^{\ssc DH} \}$,
the Heisenberg reference state is then taken as $\left| 00 \rra$ with then
$\left| \Psi \rra = U_\Psi \left| 00 \rra$. A generic $U_\Psi$ is given by
$( U_{\!\ssc A} \otimes U_{\!\ssc B} ) U_q U_{\!\ssc I3}$
where
\bea &&
U_{\!\ssc I3}=\left( \begin{array}{cc}
1 & 0 \\
0 & U_{\!\ssc 3} 
\end{array}  \right),
\qquad
U_{\!\ssc A} =\left(\begin{array}{cc}
c_{\!\ssc A} & - \bar{s}_{\!\ssc A} \\
s_{\!\ssc A} & \bar{c}_{\!\ssc A} \\
\end{array}  \right),
\qquad
U_{\!\ssc B} =\left(\begin{array}{cc}
c_{\!\ssc B} & - \bar{s}_{\!\ssc B} \\
s_{\!\ssc B} & \bar{c}_{\!\ssc B} \\
\end{array}  \right),
\sea\hspace*{1in}
U_q= \left( \begin{array}{ccc}
e^{\frac{-i\zeta}{2}} q_+ &  0  & -e^{\frac{-i\zeta}{2}} q_-  \\
0    &   I_{\!\ssc 2}  &  0	\\
e^{\frac{i\zeta}{2}} q_- &  0  &  e^{\frac{i\zeta}{2}} q_+ 
\end{array}  \right),
\nonumber\eea
and $U_{\!\ssc 3}$ any $SU(3)$ matrix. One can check that $U_\Psi$, or simply 
$( U_{\!\ssc A} \otimes U_{\!\ssc B} ) U_q$, gives exactly $\left| \Psi \rra$ 
as the first column vector. Adding back an overall phase factor to $U_\Psi$ 
by multiplying the factor to $U_\Psi$ and allowing a generic $U_{\!\ssc 3}$ 
with eight real parameters, one gets back the fifteen-parameter $SU(4)$ matrix. 
So long as the Deutsch-Hayden matrix values for observables at the particular
instances of time are concerned, we can take $( U_{\!\ssc A} \otimes U_{\!\ssc B} ) U_q$ 
as our choice of $U_\Psi$. Anyway, we see a $U_\Psi$ has exactly the information 
of the six real parameters that describe the state $\left| \Psi \rra$. For this
particular $U_\Psi$, we do not have all the matrix elements expressed in terms
of the four complex coordinates of $\left| \Psi \rra$ though.   

{The explicit form of the full expression of
$U_\Psi$ as $( U_{\!\ssc A} \otimes U_{\!\ssc B} ) U_q$ 
is still quite tedious.} Besides, with the lastly mentioned undesirable properties, 
it is not particularly illustrative to look at. However, so long as the locality issue 
is concerned, it is illustrative enough to look at the special case of $U_\psi=U_q$, 
where $\left| \psi \rra$ is the entangled state 
$e^{\frac{-i\zeta}{2}} q_+ \left| 00 \rra + e^{\frac{i\zeta}{2}} q_- \left| 11 \rra$ 
generalizing a Bell state. We have then the explicit Deutsch-Hayden matrix 
values of the local basic observables obtained as
\bea &&
[\tilde{\sigma}_{\!\ssc 1A}]_\psi^{\ssc DH}
=\left( \begin{array}{cccc}
0 & \bar{z}_{11} & \bar{z}_{00} & 0 \\
{z}_{11} & 0 & 0 & \bar{z}_{00} \\
{z}_{00} & 0 & 0 & -\bar{z}_{11} \\
0 & {z}_{00} & - {z}_{11} & 0 
\end{array}  \right)
=\left( \begin{array}{cccc}
0 & e^{\frac{-i\zeta}{2}} \sqrt{\frac{1-r}{2}} & e^{\frac{i\zeta}{2}} \sqrt{\frac{1+r}{2}} & 0 \\
e^{\frac{i\zeta}{2}} \sqrt{\frac{1-r}{2}} & 0 & 0 & e^{\frac{i\zeta}{2}} \sqrt{\frac{1+r}{2}} \\
e^{\frac{-i\zeta}{2}} \sqrt{\frac{1+r}{2}} & 0 & 0 & -e^{\frac{-i\zeta}{2}} \sqrt{\frac{1-r}{2}} \\
0 & e^{\frac{-i\zeta}{2}} \sqrt{\frac{1+r}{2}} & - e^{\frac{i\zeta}{2}} \sqrt{\frac{1-r}{2}} & 0
\end{array}  \right),
\sea
[\tilde{\sigma}_{\!\ssc 1B}]_\psi^{\ssc DH}
=\left( \begin{array}{cccc}
0 & \bar{z}_{00} & \bar{z}_{11} & 0 \\
{z}_{00} & 0 & 0 & - \bar{z}_{11} \\
{z}_{11} & 0 & 0 & {z}_{00} \\
0 & -{z}_{11} & \bar{z}_{00} & 0 
\end{array}  \right)
=\left( \begin{array}{cccc}
0 & e^{\frac{i\zeta}{2}} \sqrt{\frac{1+r}{2}} & e^{\frac{-i\zeta}{2}} \sqrt{\frac{1-r}{2}} & 0 \\
e^{\frac{-i\zeta}{2}} \sqrt{\frac{1+r}{2}} & 0 & 0 & -e^{\frac{-i\zeta}{2}} \sqrt{\frac{1-r}{2}} \\
e^{\frac{i\zeta}{2}} \sqrt{\frac{1-r}{2}} & 0 & 0 & e^{\frac{-i\zeta}{2}} \sqrt{\frac{1+r}{2}} \\
0 & -e^{\frac{i\zeta}{2}} \sqrt{\frac{1-r}{2}} & e^{\frac{i\zeta}{2}} \sqrt{\frac{1+r}{2}} & 0
\end{array}  \right),
\sea 
[\tilde{\sigma}_{\!\ssc 3A}]_\psi^{\ssc DH} = [\tilde{\sigma}_{\!\ssc 3B}]_\psi^{\ssc DH}
= \left( \begin{array}{cccc}
|{z}_{00}|^2 - |{z}_{11}|^2 & 0 & 0 & -\bar{z}_{00} \bar{z}_{11} \\
0 & 1 & 0 & 0 \\
0 & 0 & -1 & 0 \\
-{z}_{00} {z}_{11} & 0 & 0 & |{z}_{11}|^2 -|{z}_{00}|^2 
\end{array}  \right)
= \left( \begin{array}{cccc}
r & 0 & 0 & -\sqrt{1-r^2} \\
0 & 1 & 0 & 0 \\
0 & 0 & -1 & 0 \\
-\sqrt{1-r^2} & 0 & 0 & -r
\end{array}  \right),
\sea 
\eea
where we have used ${z}_{01}={z}_{10}=0$. Note that the reduced density matrices
for the qubits do not contain any information about $\zeta$, though they do have
$r$ showing up as the purity parameter showing any value of $r$ other than $1$
they are not pure states.

For our noncommutative values, we have
\[
[\zb]_\Psi = \{ f_{\!\ssc\zb}|_{\!\ssc\Psi}; V_{\!\ssc\zb}^{\ssc 00}|_{\!\ssc\Psi},
V_{\!\ssc\zb}^{\ssc 01}|_{\!\ssc\Psi}, V_{\!\ssc\zb}^{\ssc 10}|_{\!\ssc\Psi},
 V_{\!\ssc\zb}^{\ssc 11}|_{\!\ssc\Psi}\} \;.
\]
The simple results for the special case of $\left| \Psi \rra$ reduced
to $\left| \psi \rra$ are given for comparison.
\bea&&
f_{\!\tilde{\sigma}_{\!\ssc 1A}} = 0 \;,
\qquad
V_{\!\tilde{\sigma}_{\!\ssc 1A}}^{\ssc 00} = - f_{\!\tilde{\sigma}_{\!\ssc 1A}} \bar{z}_{00} =0 \;,
\qquad
V_{\!\tilde{\sigma}_{\!\ssc 1A}}^{\ssc 01} = \bar{z}_{11} -0 = e^{\frac{-i\zeta}{2}} \sqrt{\frac{1-r}{2}} \;,
\sea
V_{\!\tilde{\sigma}_{\!\ssc 1A}}^{\ssc 10}= \bar{z}_{00} -0 = e^{\frac{i\zeta}{2}} \sqrt{\frac{1+r}{2}} \;,
\qquad
V_{\!\tilde{\sigma}_{\!\ssc 1A}}^{\ssc 11} = - f_{\!\tilde{\sigma}_{\!\ssc 1A}} \bar{z}_{11} =0 \;,
\sea
f_{\!\tilde{\sigma}_{\!\ssc 1B}} = 0 \;,
\qquad
V_{\!\tilde{\sigma}_{\!\ssc 1B}}^{\ssc 00} = - f_{\!\tilde{\sigma}_{\!\ssc 1B}} \bar{z}_{00} =0 \;,
\qquad
V_{\!\tilde{\sigma}_{\!\ssc 1B}}^{\ssc 01} = \bar{z}_{00} -0 = e^{\frac{i\zeta}{2}} \sqrt{\frac{1+r}{2}} \;,
\sea
V_{\!\tilde{\sigma}_{\!\ssc 1B}}^{\ssc 10}= \bar{z}_{11} -0 = e^{\frac{-i\zeta}{2}} \sqrt{\frac{1-r}{2}} \;,
\qquad
V_{\!\tilde{\sigma}_{\!\ssc 1B}}^{\ssc 11} = - f_{\!\tilde{\sigma}_{\!\ssc 1B}} \bar{z}_{11} =0 \;,
\sea
f_{\!\tilde{\sigma}_{\!\ssc 3A}} = |{z}_{00}|^2 - |{z}_{11}|^2 = r \;,
\qquad
V_{\!\tilde{\sigma}_{\!\ssc 3A}}^{\ssc 00} = (1- f_{\!\tilde{\sigma}_{\!\ssc 3A}})\bar{z}_{00}
 = (1-r) e^{\frac{i\zeta}{2}} \sqrt{\frac{1+r}{2}} \;,
\sea
V_{\!\tilde{\sigma}_{\!\ssc 3A}}^{\ssc 01} = 0 \;,
\qquad
V_{\!\tilde{\sigma}_{\!\ssc 3A}}^{\ssc 10} = 0 \;,
\qquad
V_{\!\tilde{\sigma}_{\!\ssc 3A}}^{\ssc 11} = - (1+ f_{\!\tilde{\sigma}_{\!\ssc 3A}})\bar{z}_{11}
=- (1+r)e^{\frac{-i\zeta}{2}} \sqrt{\frac{1-r}{2}} \;,
\eea
and again $[\tilde{\sigma}_{\!\ssc 3A}]_{\phi}=[\tilde{\sigma}_{\!\ssc 3B}]_{\phi}$.
We note in particular that for $r=1$, $\left| \psi \rra$ reduces to the product state 
$e^{\frac{-i\zeta}{2}} \left| 00 \rra$ (as $q_+=1$ and $q_-=0$) for which we have
$[\tilde{\sigma}_{\!\ssc 1A}]_{\phi (r=1)} = \{ 0; 0, 0, e^{\frac{i\zeta}{2}}, 0\}$,
$[\tilde{\sigma}_{\!\ssc 1B}]_{\phi (r=1)} = \{ 0; 0, e^{\frac{i\zeta}{2}}, 0, 0\}$,
and $[\tilde{\sigma}_{\!\ssc 3A}]_{\phi (r=1)} = [\tilde{\sigma}_{\!\ssc 3B}]_{\phi (r=1)}
	= \{ 1; 0, 0, 0, 0\}$. The last result is the consequence that the state
is an eigenstate of $\tilde{\sigma}_{\!\ssc 3A}$ and $\tilde{\sigma}_{\!\ssc 3B}$
of eigenvalue, and hence expectation value, $1$. As mentioned in the main text,
for an eigenstate of an observable, the noncommutative value becomes essentially 
a commutative value. $e^{\frac{i\zeta}{2}}$, in the case, is just an overall phase
factor, the conjugate of that for the state. We can drop it and take
$[\tilde{\sigma}_{\!\ssc 1A}]_{\phi (r=1)} = \{ 0; 0, 0, 1, 0\}$ and
$[\tilde{\sigma}_{\!\ssc 1B}]_{\phi (r=1)} = \{ 0; 0, 1, 0, 0\}$.

\section*{Appendix B:  Noncommutative Values of 
Local Basic Observables for the Two-particle System (The exact EPR case)}

Let us look at a different example of the noncommutative value picture for the 
local basic observables of a system of two quantum particles concerning the  
locality issue of the exact EPR state. Noncommutative values for the position
and momentum operators under a Schr\"odinger wavefunction representation
have been explicitly given in Ref.\cite{093}. Here, we adapt that to the 
two-particle wavefunction $\phi(x_{\ssc 1},x_{\ssc 2})$.  For convenience, we
take the particles to be of equal mass $m$. Take the entangled state of fixed total 
momentum $p$ ($\hbar$ taken as unity), and a separation between the particles
given by $r= vt$ increasing linearly with time.  $p$ and $r$ should then by
eigenvalues of the dynamics observables $\hat{P}$ and $\hat{R}$ to which the
system maintains in simultaneous eigenstates. Note that the observables are
parts of the canonical pairs of observables for the center of mass and relative
motion degrees of freedom with
\bea &&
\hat{X} = \frac{\hat{X}_{\!\ssc 1} + \hat{X}_{\!\ssc 2}}{2}  \;,
\qquad 
\hat{P} = \hat{P}_{\!\ssc 1} + \hat{P}_{\!\ssc 2} \;,
\sea
\hat{R} = \hat{X}_{\!\ssc 1} - \hat{X}_{\!\ssc 2} \;,
\qquad \label{cv}
\hat{Q} = \frac{\hat{P}_{\!\ssc 1} - \hat{P}_{\!\ssc 2}}{2} \;,
\eea
with $[\hat{X} ,\hat{P} ] = i=[\hat{R} ,\hat{Q}]$ as the only nonzero commutators
among them. The wavefunction of the EPR state is then written as 
\bea
= \delta(r-r_o) e^{ipx}  
=  \delta(x_{\ssc 1}-x_{\ssc 2}-r_o) \,  e^{\frac{ip(x_{\ssc 1}+x_{\ssc 2})}{2}}  \;,
\eea
where $x=\frac{x_{\ssc 1}+x_{\ssc 2}}{2}$, and $r= x_{\ssc 1}-x_{\ssc 2}$.
The state is, of course, not an eigenstate of any of the  local basic observables
as the position and momentum of the individual particles.  The entanglement
is seen in the fact that the wavefunction $\phi(x_{\ssc 1},x_{\ssc 2})$ cannot
be written in terms of a product of a $\phi_{\ssc 1}(x_{\ssc 1})$ and 
a $\phi_{\ssc 2}(x_{\ssc 2})$ as wavefunctions for states of the individual
particles. In terms of the degrees of freedom for the canonical observables 
of Eq.(\ref{cv}), {\em i.e.} in terms of real variables $r$ and $x$, however,
there is no entanglement.

%We have a time evolution given by the Hamiltonian 
%$\hat{H}=\frac{ \hat{P}^2}{4m} + \frac{ \hat{Q}^2}{2m}$, hence
%$\hat{U}(t) = e^{\frac{it}{4m} (\partial_{x}^2+2\partial_r^2)$.
 
The noncommutative values for $\hat{P}$ and $\hat{R}$, assuming the matrix 
elements, are given by  $[\hat{P}]_\phi = \{p,;V_{\!\ssc \hat{P}}\}$ and 
$[\hat{R}]_\phi = \{vt; V_{\!\ssc \hat{R}}\}$ where \cite{093}
\bea &&
V_{\!\ssc \hat{P}} \equiv \delta_\phi f_{\!\ssc \hat{P}} = 0 \;,
\sea
V_{\!\ssc \hat{R}} \equiv \delta_\phi f_{\!\ssc \hat{R}} = 0 \;.
\eea
The last equations express the collection of `coordinates' derivatives of the
corresponding expectation value function(al)s as the functional derivatives. The 
vanishing results are a consequence of $\phi$ being an eigenstate of the observables. 
We further take the expectation values of all the position observables at $t=0$
and that of the momentum observable $\hat{Q}$ to be zero. We have then
$[\hat{X}]_\phi = \{\bar{x}; V_{\!\ssc \hat{X}}\}$ and 
$[\hat{Q}]_\phi = \{\bar{q}; V_{\!\ssc \hat{Q}}\}$ where 
\bea &&
V_{\!\ssc \hat{X}} \equiv \delta_\phi f_{\!\ssc \hat{X}} 
= e^{-ipx}  \delta(r-r_o) \,(x-\bar{x}) \;,
%= \delta(x_{\ssc 1}-x_{\ssc 2}-vt) \,\frac{x_{\ssc 1}+x_{\ssc 2}}{2} e^{\frac{ip(x_{\ssc 1}+x_{\ssc 2})}{2}}  \;,
\sea
V_{\!\ssc \hat{Q}} \equiv \delta_\phi f_{\!\ssc \hat{Q}} 
= e^{-ipx} (i\partial_r -\bar{q}) \delta(r-r_o) \;.
\eea
Note that $\bar{x}$ and $\bar{q}$ denote the expectation values of the 
observables. We expect their values to be zero. Yet, it serves our purpose here
better to simply leave them in the direct formal integrals as
\bea &&
\bar{x}= \int\!\!dx\; x |e^{ipx}|^2 \;,
\sea
\bar{q} = \int\!\!dr\;  \delta(r-r_o) (-i\partial_r) \delta(r-r_o) \;,
\eea 
to be used below. All the expressions involving the noncommutative values we
have given so far factorize into products of functions of $x$ and functions of $r$.
When rewritten in terms of variables $x_{\ssc 1}$ and $x_{\ssc 2}$, 
$V_{\!\ssc \hat{X}}$ and $V_{\!\ssc \hat{Q}}$ do not factorize. That indicates
the nature of the state as being an entangled one in the corresponding degrees
of freedom, {\em i.e.} those of the individual particles.

For the local basic observables, the noncommutative values can be obtained as
$[\hat{X}_{\!\ssc 1}]_\phi = \{\bar{x}_{\!\ssc 1}; V_{\!\ssc \hat{X}_{\ssc 1}}\}$, 
$[\hat{X}_{\!\ssc 2}]_\phi = \{\bar{x}_{\!\ssc 2}; V_{\!\ssc \hat{X}_{\ssc 2}}\}$,
$[\hat{P}_{\!\ssc 1}]_\phi = \{\bar{x}_{\!\ssc 1}; V_{\!\ssc \hat{P}_{\ssc 1}}\}$, and
$[\hat{P}_{\!\ssc 2}]_\phi = \{\bar{x}_{\!\ssc 2}; V_{\!\ssc \hat{P}_{\ssc 2}}\}$, with
the expectation values formally as
\bea &&
\bar{x}_{\!\ssc 1} = \int\!\!dx dr\; {x}_{\!\ssc 1} \, |e^{ipx}|^2 \delta^2(r-r_o)
= \bar{x}+ \frac{r_o}{2} \; ,
\sea
\bar{x}_{\!\ssc 2} = \int\!\!dx dr\; {x}_{\!\ssc 2} \, |e^{ipx}|^2 \delta^2(r-r_o)
= \bar{x}- \frac{r_o}{2} \; ,
\sea
\bar{p}_{\!\ssc 1} =   e^{-ipx} \delta(r-r_o) (-i\partial_{x_{\!\ssc 1}})  e^{ipx} \delta(r-r_o)
= \frac{p}{2}  +\int\!\! dr\; \delta(r-r_o) (-i\partial_r) \delta(r-r_o) \;,
\sea
\bar{p}_{\!\ssc 1} =   e^{-ipx} \delta(r-r_o) (-i\partial_{x_{\!\ssc 1}})  e^{ipx} \delta(r-r_o)
= \frac{p}{2}  - \int\!\! dr\; \delta(r-r_o) (-i\partial_r) \delta(r-r_o) \;,
\eea
and
\bea &&
V_{\!\ssc \hat{X}_{\ssc1}} = e^{-ipx}  \delta(r-r_o) \,(x_{\!\ssc 1} -\bar{x}_{\!\ssc 1} )  \;,
\sea
V_{\!\ssc \hat{X}_{\ssc 2}} = e^{-ipx}  \delta(r-r_o) \,(x_{\!\ssc 2} -\bar{x}_{\!\ssc 2} ) 
= (x_{\!\ssc 2} -\bar{x}_{\!\ssc 2}) \delta(x_{\ssc 1}-x_{\ssc 2}-r_o) \, e^{\frac{-ip(x_{\ssc 1}+x_{\ssc 2})}{2}} \;,
%= e^{-ipx}  \delta(r-vt)  \left[ x-\bar{x} - \frac{r-vt}{2} \right] \;,
\sea
V_{\!\ssc \hat{P}_{\ssc 1}} = (i \partial_{x_{\!\ssc 1}}- \bar{p}_{\!\ssc 1}) e^{-ipx}  \delta(r-r_o)
= e^{-ipx} \left( \frac{p}{2}  + i \partial_r - \bar{p}_{\!\ssc 1} \right)   \delta(r-r_o) \;,
\sea
V_{\!\ssc \hat{P}_{\ssc 2}} =  (i \partial_{x_{\!\ssc 2}}- \bar{p}_{\!\ssc 2}) e^{-ipx}  \delta(r-r_o)
= e^{-ipx} \left( \frac{p}{2}  - i \partial_r - \bar{p}_{\!\ssc 2} \right)   \delta(r-r_o) \;.
\eea
The results above clearly demonstrate the linearity of the noncommutative values
giving 
\bea &&
[\hat{X}]_\phi  = \frac{[\hat{X}_{\!\ssc 1}]_\phi  + [\hat{X}_{\!\ssc 2}]_\phi }{2}  \;,
\qquad 
[\hat{P}]_\phi  = [\hat{P}_{\!\ssc 1}]_\phi  + [\hat{P}_{\!\ssc 2}]_\phi  \;,
\sea
[\hat{R}]_\phi  = [\hat{X}_{\!\ssc 1} ]_\phi - [\hat{X}_{\!\ssc 2}]_\phi  \;,
\qquad 
[\hat{Q}]_\phi = \frac{[\hat{P}_{\!\ssc 1}]_\phi - [\hat{P}_{\!\ssc 2}]_\phi }{2} \;.
\eea
%\bea &&
%V_{\!\ssc \hat{X}} = \frac{V_{\!\ssc \hat{X}_{\!\ssc 1}} +V_{\!\ssc  \hat{X}_{\!\ssc 2}}}{2}  \;,
%\qquad 
%V_{\!\ssc \hat{P}} = V_{\!\ssc \hat{P}_{\!\ssc 1}} + V_{\!\ssc \hat{P}_{\!\ssc 2}} \;,
%\sea
%V_{\!\ssc \hat{R}} = V_{\!\ssc \hat{X}_{\!\ssc 1}} - V_{\!\ssc \hat{X}_{\!\ssc 2}} \;,
%\qquad \label{cv}
%V_{\!\ssc \hat{Q}} = \frac{V_{\!\ssc \hat{P}_{\!\ssc 1}} - V_{\!\ssc \hat{P}_{\!\ssc 2}}}{2} \;,
%\eea
We can further express the noncommutative values for the local basic 
observables as 
\bea &&
[\hat{X}_{\!\ssc 1}]_\phi
= \left\{  \bar{x}+ \frac{r_o}{2} ;  \left(x_{\!\ssc 1} -\bar{x} - \frac{r_o}{2}\right) \delta(x_{\ssc 1}-x_{\ssc 2}-r_o) \, 
    e^{\frac{-ip(x_{\ssc 1}+x_{\ssc 2})}{2}} \right\} \;,
\sea
[\hat{X}_{\!\ssc 2}]_\phi
= \left\{  \bar{x}- \frac{r_o}{2} ;  \left(x_{\!\ssc 2} -\bar{x} + \frac{r_o}{2} \right) \delta(x_{\ssc 1}-x_{\ssc 2}-r_o) \, 
    e^{\frac{-ip(x_{\ssc 1}+x_{\ssc 2})}{2}}  \right\} \;,
\sea
[\hat{P}_{\!\ssc 1}]_\phi
= \left\{  \frac{p}{2} + \bar{q};  e^{\frac{-ip(x_{\ssc 1}+x_{\ssc 2})}{2}} (i \partial_{x_{\!\ssc 1}} -\bar{q}) 
   \delta(x_{\ssc 1}-x_{\ssc 2}-r_o)    \right\} \;,
\sea
[\hat{P}_{\!\ssc 2}]_\phi
= \left\{  \frac{p}{2} - \bar{q};  e^{\frac{-ip(x_{\ssc 1}+x_{\ssc 2})}{2}} (i \partial_{x_{\!\ssc 2}} +\bar{q}) 
   \delta(x_{\ssc 1}-x_{\ssc 2}-r_o)    \right\} \;,
\eea
showing explicitly how they encode the information that would be seen as nonlocal 
in the language of the states. $V_{\!\ssc \hat{X}_{\ssc1}}$ of $[\hat{X}_{\!\ssc 1}]_\phi$, 
for example, involves the $x_{\ssc 2}$ values and the nonfactorizable nature 
of it indicates the entanglement. Note that the nonvanishing $V_{\!\ssc \hat{X}_{\ssc 1}}$,
$V_{\!\ssc \hat{X}_{\ssc 2}}$, $V_{\!\ssc \hat{P}_{\ssc 1}}$, and $V_{\!\ssc \hat{P}_{\ssc 2}}$
parts of the noncommutative values indicate that the state is not an eigenstate 
of the observables. The Heisenberg uncertainty of an observable $\hat{A}$ 
is simply given by the integral of $|V_{\!\ssc \hat{A}}|^2$. The cancellation
between $V_{\!\ssc \hat{X}_{\ssc 1}}$ and $V_{\!\ssc \hat{X}_{\ssc 2}}$ 
in $V_{\!\ssc \hat{R}}$ and that of $V_{\!\ssc \hat{P}_{\ssc 1}}$ and 
$-V_{\!\ssc \hat{P}_{\ssc 2}}$ in $V_{\!\ssc \hat{P}}$ is the statement
that the state is an eigenstate of  $\hat{R}$ and $\hat{P}$ with the exact
correlations between the pairs of observables involved. Explicitly, we have
\bea &&
V_{\!\ssc \hat{R}} = V_{\!\ssc \hat{X}_{\ssc 1}} - V_{\!\ssc \hat{X}_{\ssc 2}}
= (x_{\ssc 1}-x_{\ssc 2}-r_o)  \delta(x_{\ssc 1}-x_{\ssc 2}-r_o) \,    e^{\frac{-ip(x_{\ssc 1}+x_{\ssc 2})}{2}} =0 \;,
\sea
V_{\!\ssc \hat{P}} = V_{\!\ssc \hat{P}_{\ssc 1}} + V_{\!\ssc \hat{P}_{\ssc 2}}
= e^{\frac{-ip(x_{\ssc 1}+x_{\ssc 2})}{2}} i (\partial_{x_{\!\ssc 1}} -\partial_{x_{\!\ssc 2}})  \delta(x_{\ssc 1}-x_{\ssc 2}-r_o) =0 \;.
\eea

%%%%%%%%%%%%%%%%%%%%%%%%%%%%%%%%

\section*{Acknowledgment:}
The work has been partially supported by research grant
number   111-2112-M-008-029
of the MOST of Taiwan, research grant 
number   112-2112-M-008-019 of the NSTC of Taiwan, and a Visiting Professorship
at Korea Institute for Advanced Study.

The study is purely theoretical, and no data is involved.
The author declares that there is no conflict of interest.

\end{document}